\documentclass[journal]{IEEEtran}
\ifCLASSINFOpdf
\else
\fi
\usepackage{amsmath}
\usepackage{pgfplots}
\pgfplotsset{compat=1.16}
\usepackage{tikz}

\begin{document}

\title{Linear Channel Estimation Based on a Low-Bandwidth Observation Channel with Unknown Response}

\author{\IEEEauthorblockN{{Juan I. Bonetti}\IEEEauthorrefmark{1},
    {James Kunst\IEEEauthorrefmark{1}, Dami\'an A. Morero\IEEEauthorrefmark{2}, and Mario R. Hueda}\IEEEauthorrefmark{2}}\\
  \IEEEauthorblockA {
    \IEEEauthorrefmark{1} Fundaci\'on Fulgor - Romagosa 518 - C\'ordoba (5000) - Argentina \\
    \IEEEauthorrefmark{2} Laboratorio de Comunicaciones Digitales - Universidad Nacional de  C\'ordoba\\
    Av. V\'elez Sarsfield 1611 - C\'ordoba (X5016GCA) - Argentina\\
    Email: juan.bonetti@ib.edu.ar}}


\maketitle

\begin{abstract}
  We propose a novel system identification technique, based on a
  least-mean square algorithm, allowing for the estimation of a linear
  channel by using an unknown-response measurement channel. The key of
  the technique is a memoryless nonlinear function working as
  uncoupling block between the estimated and observation channels,
  conforming a Wiener-Hammerstein scheme. We prove that this
  estimation, only differing from the actual channel response by a
  scaling factor and a temporal shift, does not depend on the
  observation channel bandwidth. As a consequence, this technique enables the usage of low-cost measurement devices as feedback channel. We present numerical examples
  of the method, supporting the proposal and displaying excellent results.
\end{abstract}


%
\IEEEpeerreviewmaketitle

\section{Introduction}
System identification, also known as channel estimation, refers to the
different techniques used to estimate the response of a system
starting from measured data of its input and
output~\cite{pintelon2012system,keesman2011system}. In a few words,
given certain \textit{model} for the system, a mathematical function
relating the input and output signals, system identification consists
in finding the parameters of that function that best fit the actual
behavior of the channel. In many situations, the first approach to the
estimation of an unknown-response system is performed by the
optimization of a linear model. In these cases, system identification
is carried out by means of an algorithm able to find the linear
function that best emulates the actual response of the
system. Moreover, this function commonly provides a suitable and
simple model of the channel.

In this work we focus in the \textit{linear estimation} case, for
which the channel of interest (channel to be estimated) is strictly
linear. We consider a particular problem associated to the linear
estimation: the measurement of the output through some other
unknown-response system, also assumed to be linear. As the measurement
system requires to be estimated, this task becomes a mathematical
problem: how to perform the system identification of a chain of two
unknown linear systems. As we explain in the next section, the
decomposition of the different channels of the chain is not possible
from a mathematical point of view. However, we show that if a
nonlinear function is added between the two channels, it will work as
an effective \textit{uncoupling block} between them, allowing for the
estimation of both systems. In general, we show that the channel
responses can be estimated unless two coefficients: one related to a
scale factor and other associated to a temporal-shift. If the
nonlinear function, expressed as a polynomial function, has at least
one even term and one odd term, the scale factor is always the
unity. In addition, if both channels are causal and \textit{not delayed}\footnote{The impulse response of the linear system has a nonzero instantaneous part.}, the
coefficient of temporal-shift is zero.

The mathematical proofs provided in this work are only valid for
scalar-real signals, but can be used as the starting point for a
further analysis, including vectorial and complex input-output
signals. Our main goal is to illustrate not only that the simultaneous
estimation of the channel of interest and the measurement system is
possible with a nonlinear function working as uncoupling device, but
also an additional advantage of this scheme: the possibility of using
ultra-low bandwidth measurement channels. This feature is commonly associated to low-cost measure devices, which are naturally expected to pose a limited frequency response and a low sample rate. A clear example is found in the coherent optical systems, where a high-performance coherent detector is usually required in the feedback channel; results derived in this work show that the coherent detector can be replaced with a low-cost nonlinear device such as a common photodiode. The numerical results
presented in the last section are focused in proving this striking
feature of the proposed method.

\section{System Identification: the Observation Channel Problem}
The digital model of a simple linear channel relates two scalar
time-discrete signals, the \textit{input sequence} $x[n]$ with the
\textit{output sequence} $y[n]$, through the sum given by
\begin{equation}
    y[n] = \sum_m L[m] x[n-m] = \sum_m L[n-m] x[m],
    \label{firsteq}
\end{equation}
where $L[m]$ are real coefficients defining the response of the
channel of interest and, in the general case, $m$ varies over all the integers. An
useful alternative representation of this simple linear system is
given by its frequency-domain version, resulting from the Fourier
transformation of the signals, defined as
\begin{equation}
  \left\lbrace \begin{array}{l} \tilde{f}(\Omega) = \sum_n f[n]e^{-\dot{\j} \Omega n}\\ f[n] = \frac{1}{2\pi}\int_{-\pi}^{\pi}\tilde{f}(\Omega)e^{\dot{\j} \Omega n} \,d\Omega.\end{array}\right.
\end{equation}
By applying this transformation on Eq.~\ref{firsteq} we obtain
\begin{equation}
  \tilde{y}(\Omega) = \tilde{L}(\Omega) \tilde{x}(\Omega).
    \label{freq_sys}
\end{equation}
Either Eq.~\ref{firsteq} or Eq.~\ref{freq_sys} can be used to obtain
the output sequence from the input sequence. However, sometimes we are
interested in knowing the response of the channel given the
input-output sequence more than the output. This problem consists in
the estimation of coefficients $L[m]$ starting from the sequences
$x[n]$ and $y[n]$, and it is a particular case of the area of study
known as \textit{system identification}~\cite{pintelon2012system}.

\begin{figure}
    \centering
    \begin{tikzpicture}
        \draw[thick,->] (0,3)--(1.5,3);
        \node at (0.75,3.25) {$x[n]$};
        \draw[thick] (1.5,3.5)rectangle(3,2.5);
        \node at (2.25,3) {$L[m]$};
         \node at (2.25,3.75) {\small{channel of interest}};
        \draw[thick,->] (3,3)--(4.5,3);
        \node at (3.75,3.25) {$y[n]$};
        \draw[thick] (1.5,2)rectangle(3,1.5);
        \node at (2.25,1.75) {LMS};
        \draw[thick] (1.5,1)rectangle(3,0);
        \node at (2.25,0.5) {$\hat{L}[m]$};
        \node at (2.25,-0.25) {\small{estimated channel}};
        \draw[thick,->] (3,0.5)--(4.5,0.5);
        \node at (3.75,0.25) {$\hat{y}[n]$};
        \draw[thick] (0.75,3)--(0.75,0.5);
        \draw[thick,->] (0.75,0.5)--(1.5,0.5);
        \draw[thick,->] (0.75,1.75)--(1.5,1.75);
        \draw[thick,->] (2.25,1.5)--(2.25,1);
        \draw[thick] (4.25,1.75)circle(0.15);
        \node at (4.25,1.7) {-};
        \draw[thick,->] (4.25,3)--(4.25,1.9);
        \draw[thick,->] (4.25,0.5)--(4.25,1.6);
        \draw[thick,->] (4.1,1.75)--(3,1.75);
        \node at (3.6,2) {$e_y[n]$};
    \end{tikzpicture}
    \caption{Least mean square (LMS) algorithm for a simple linear
      system. The algorithm minimizes the mean square of the error
      signal, $E\lbrace e^2_y[n]\rbrace$, obtaining an estimation of
      the actual channel, given by $\hat{L}[m] = L[m]$.}
    \label{fig01}
\end{figure}
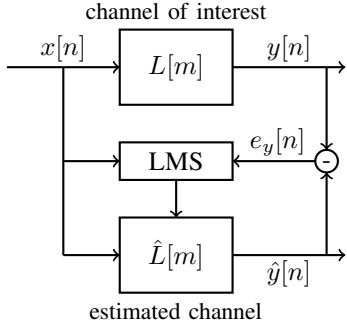

Figure~\ref{fig01} shows the basic scheme of an usual algorithm for
the system identification of linear systems: the least-mean squares
(LMS) algorithm~\cite{widrow1985adaptive}. The estimated coefficients
$\hat{L}[m]$, whose initial values are commonly initialized according
to a rough estimation of the actual channel $L[m]$, produce the
\textit{estimated output sequence},
\begin{equation}
    \hat{y}[n] = \sum_m \hat{L}[m] x[n-m],
\end{equation}
that is compared with the actual output sequence to calculate the
\textit{error signal}
\begin{equation}
    e_y[n] = \hat{y}[n] - y[n].
\end{equation}
The LMS algorithm allows for the minimizing of the mean squared error
(MSE)
\begin{equation}
  \mathcal{\hat E} = E\lbrace e_y^2[n] \rbrace,
    \label{errorsquare}
\end{equation}
where $E\lbrace .\rbrace$ is the expectation operator. The MSE
\eqref{errorsquare} is assumed to be a continuous and differentiable
function of the estimated coefficients $\hat{L}[m]$. Moreover, the
algorithm allows for the calculation of the coefficients that minimize
Eq.~\ref{errorsquare},
\begin{equation}
  \hat{L}[m] = \underset{\hat{L}[m]}{\mathrm{argmin}}\left[
    \mathcal{\hat E}\right].
\end{equation}
It is easy to prove that for the scheme of Fig.~\ref{fig01} the
minimum MSE is zero and it is achieved when $\hat{L}[m] = L[m]$.

\subsection{Linear Observation Channel}
\begin{figure}
    \centering
    \begin{tikzpicture}
        \draw[thick,->] (0,3)--(1.5,3);
        \node at (0.75,3.25) {$x[n]$};
        \draw[thick] (1.5,3.5)rectangle(3,2.5);
        \node at (2.25,3) {$L[m]$};
        \node at (2.25,3.75) {\small{channel of interest}};
        \draw[thick,->] (3,3)--(4.5,3);
        \node at (3.75,3.25) {$y[n]$};
        \draw[thick] (4.5,3.5)rectangle(6,2.5);
        \node at (5.25,3) {$O[m]$};
        \node at (5.25,3.75) {\small{observation channel}};
        \draw[thick,->] (6,3)--(7.5,3);
        \node at (6.75,3.25) {$z[n]$};
        \draw[thick] (1.5,2)rectangle(6,1.5);
        \node at (3.75,1.75) {LMS};
        \draw[thick] (1.5,1)rectangle(3,0);
        \node at (2.25,0.5) {$\hat{L}[m]$};
        \node at (2.25,-0.25) {\small{estimated channel}};
        \draw[thick,->] (3,0.5)--(4.5,0.5);
        \node at (3.75,0.25) {$\hat{y}[n]$};
        \draw[thick] (4.5,1)rectangle(6,0);
        \node at (5.25,0.5) {$\hat{O}[m]$};
        \node at (5.25,-0.25) {\small{ estimated feedback}};
        \draw[thick,->] (5.25,1.5)--(5.25,1);
        \draw[thick,->] (6,0.5)--(7.5,0.5);
        \node at (6.75,0.25) {$\hat{z}[n]$};
        \draw[thick] (0.75,3)--(0.75,0.5);
        \draw[thick,->] (0.75,0.5)--(1.5,0.5);
        \draw[thick,->] (0.75,1.75)--(1.5,1.75);
        \draw[thick,->] (2.25,1.5)--(2.25,1);
        \draw[thick] (7.25,1.75)circle(0.15);
        \node at (7.25,1.7) {-};
        \draw[thick,->] (7.25,3)--(7.25,1.9);
        \draw[thick,->] (7.25,0.5)--(7.25,1.6);
        \draw[thick,->] (7.1,1.75)--(6,1.75);
        \node at (6.6,2) {$e[n]$};
    \end{tikzpicture}
    \caption{Least mean square (LMS) algorithm including the
      estimation of the observation channel $\hat{O}[m]$. Although the
      mean error square is suppressed, the estimated systems does not
      necessarily match with the actual systems. In general
      $\hat{L}[m] \neq L[m]$ and $\hat{O}[m] \neq O[m]$.}
    \label{fig02}
\end{figure}
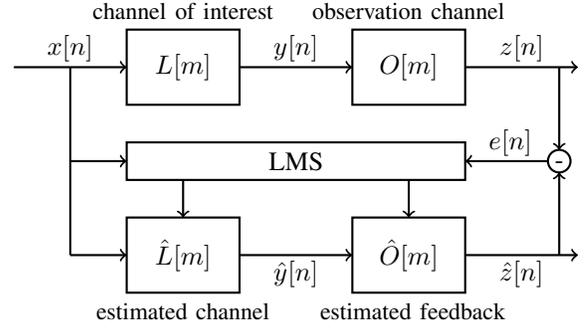

A common obstacle to implement a simple LMS algorithm, is the fact
that the output sequence is not directly accessible. An
\textit{observation channel}, such as a transductor or any
data-acquisition device, is used to obtain information about the
system output. In the simplest scheme, the observation channel is a
linear system that can be modeled by means of the coefficients $O[m]$,
as shown in Fig.~\ref{fig02}, leading to the distorted version or
indirect measurement of the output signal, the \textit{measurement
  sequence}
\begin{equation}
    z[n] = \sum_m O[m] y[n-m].
    \label{measuresignal}
\end{equation}
The most typical situation is that of the response of this observation
channel is not completely known. As a consequence, the system
identification must include its estimation, $\hat{O}[m]$, as displayed
in Fig.~\ref{fig02}. However, this scheme presents two several
limitations. In Appendix~\ref{lms_problem} we show that once the
algorithm reaches the stationary state, for which the error signal is
suppressed, the estimated systems and the actual channels satisfy
\begin{equation}
    \breve{L}(\Omega)\breve{O}(\Omega) = \tilde{L}(\Omega)\tilde{O}(\Omega),
    \label{lms_convergence}
\end{equation}
where $\breve{L}(\Omega)$ and $\breve{O}(\Omega)$ are the Fourier
transforms of $\hat{L}[m]$ and $\hat{O}[m]$, respectively. This result
suggests that the simultaneous estimation of the channel of interest and the
observation channel can not be performed, as Eq.~\ref{lms_convergence}
does not imply $\hat{L}[m] = L[m]$ or $\hat{O}[m] = O[m]$. Moreover,
there exist an infinity number of combinations of $\hat{L}[m]$ and
$\hat{O}[m]$ satisfying Eq.~\ref{lms_convergence} and being completely
uncorrelated with the actual response of the channels. On the other
hand, even for the case in which the observation channel is assumed to
be known, its bandwidth must be at least equal to that of the
$\tilde{L}(\Omega)$, in order to do not miss the information about
high-frequency components of the channel response. In many cases, this
condition seriously deepens the requirements on the observation
channel, augmenting its cost and complexity.

\section{The Nonlinear Observation Channel}
In order to solve the observation channel problems presented in the
previous section, we propose the modified scheme shown in Fig. 3. A
known memoryless nonlinear function, described by the real
coefficients $a_q$, works as an uncoupling block between the estimated
channel and the observation channel. Particularly, the nonlinear
function produces the output
\begin{equation}
    s[n] = \sum_{q=1}^{Q} a_q y^q[n],
\end{equation}
where $Q$ is the function degree and $a_q$ is assumed to be nonzero
for at least one $q>1$. This structure, conformed by the nonlinear
function between two linear channel, is known as a Wiener-Hammerstein
filter and is commonly found in the modeling of linear plants with
nonlinear-outputs sensors, such as photodiodes, preasure gauges,
thermocouplers, between others~\cite{schoukens2019nonlinear}. The coefficients $a_q$ are assumed to be known as they commonly describe the transduction principle of the measurement, based in a well-known physical law. For instance, a simple photodiode is assumed to follow a quadratic relation ($a_2 = 1$) between the optical input and the electrical output. Scaling factor and frequency response of the a non-ideal device are modeled by the unknown response $O[m]$. However, the estimation of the $a_q$ coefficients could be also performed with the LMS algorithm, but it is a topic of our current research and further investigation is required.

\begin{figure*}
    \centering
    \begin{tikzpicture}
        \draw[thick,->] (0,3)--(1.5,3);
        \node at (0.75,3.25) {$x[n]$};
        \draw[thick] (1.5,3.5)rectangle(3,2.5);
        \node at (2.25,3) {$L[m]$};
        \node at (2.25,3.75) {\small{channel of interest}};
        \draw[thick,->] (3,3)--(4.5,3);
        \node at (3.75,3.25) {$y[n]$};
        \draw[thick] (4.5,3.5)rectangle(6,2.5);
        \node at (5.25,3) {$a_q$};
        \node at (5.25,3.75) {\small{nonlinear function}};
        \node at (5.25,4) {\small{memoryless}};
        \draw[thick,->] (6,3)--(7.5,3);
        \node at (6.75,3.25) {$s[n]$};
        \draw[thick] (7.5,3.5)rectangle(9,2.5);
        \node at (8.25,3) {$O[m]$};
        \node at (8.25,3.75) {\small{observation channel}};
        \draw[thick,->] (9,3)--(10.5,3);
        \node at (9.75,3.25) {$z[n]$};
        \draw[thick] (1.5,2)rectangle(9,1.5);
        \node at (5.25,1.75) {LMS};
        \draw[thick] (1.5,1)rectangle(3,0);
        \node at (2.25,0.5) {$\hat{L}[m]$};
        \node at (2.25,-0.25) {\small{estimated channel}};
        \draw[thick,->] (3,0.5)--(4.5,0.5);
        \node at (3.75,0.25) {$\hat{y}[n]$};
        \draw[thick] (4.5,1)rectangle(6,0);
        \node at (5.25,0.5) {$a_q$};
        \draw[thick,<->] (8.25,1.5)--(8.25,1);
        \draw[thick,->] (6,0.5)--(7.5,0.5);
        \node at (6.75,0.25) {$\hat{s}[n]$};
        \draw[thick] (7.5,1)rectangle(9,0);
        \node at (8.25,0.5) {$\hat{O}[m]$};
        \node at (8.25,-0.25) {\small{estimated feedback}};
        \draw[thick,->] (9,0.5)--(10.5,0.5);
        \node at (9.75,0.25) {$\hat{z}[n]$};
        \draw[thick] (0.75,3)--(0.75,0.5);
        \draw[thick,->] (0.75,0.5)--(1.5,0.5);
        \draw[thick,->] (0.75,1.75)--(1.5,1.75);
        \draw[thick,->] (2.25,1.5)--(2.25,1);
        \draw[thick] (10.25,1.75)circle(0.15);
        \node at (10.25,1.7) {-};
        \draw[thick,->] (10.25,3)--(10.25,1.9);
        \draw[thick,->] (10.25,0.5)--(10.25,1.6);
        \draw[thick,->] (10.1,1.75)--(9,1.75);
        \node at (9.6,2) {$e[n]$};
    \end{tikzpicture}
    \caption{Basic scheme of the proposed LMS algorithm. The
      memoryless nonlinear function, described by the coefficients
      $a_k$ allows for the decoupling between the estimated channel
      $L[m]$ and the observation channel $O[m]$. In addition, we show
      that this scheme allows for an ultra low-bandwidth response of
      the measurement channel.}
    \label{fig03}
\end{figure*}
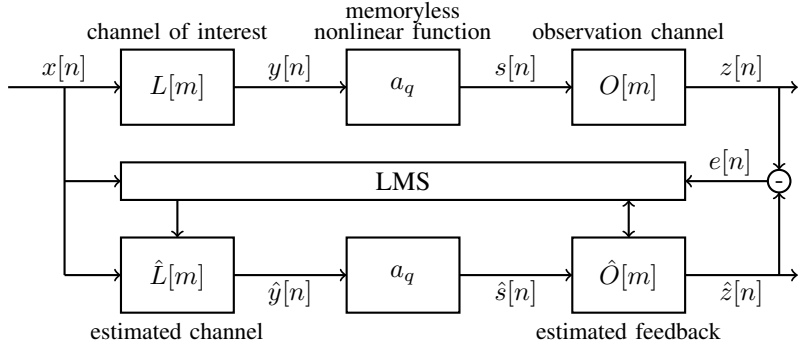

We propose the LMS algorithm to be implemented as an standard
\textit{gradient descent} algorithm, for which the cost function is
the MSE defined by 
\begin{equation}
  \mathcal{E} = E\lbrace e^2[n] \rbrace,
  \label{errorsquare2}
\end{equation}
where
\begin{equation}
    e[n] = \hat{z}[n] - z[n].
\end{equation}
Notice that the MSE \eqref{errorsquare2} is a function of both
$\hat{L}[m]$ and $\hat{O}[m]$. Thus, the adaptation of the estimation
coefficients is calculated as
\begin{equation}
  \left\lbrace \begin{array}{l} \hat{L}^{(k+1)}[m] = \hat{L}^{(k)}[m] - \beta \frac{\partial \mathcal{E}}{\partial \hat{L}[m]} \\ \hat{O}^{(k+1)}[m] = \hat{O}^{(k)}[m] - \beta \frac{\partial \mathcal{E}}{\partial \hat{O}[m]} \end{array}  \right.,
\end{equation}
where $\beta > 0$ is the \textit{adaptation step}.  In
Appendix~\ref{derivatives} we calculate the derivatives, leading to
\begin{multline}
  \hat{L}^{(k+1)}[m] = \hat{L}^{(k)}[m] - \\2\beta \left\lbrace e[n]
    \sum_r \hat{O}[r]\sum_{q=1}^Q a_q q \hat{y}^{q-1}[n-r]
    x[n-r-m]\right\rbrace
    \label{grad1}
\end{multline}
and
\begin{equation}
  \hat{O}^{(k+1)}[m] = \hat{O}^{(k)}[m] -\\ 2\beta\left\lbrace e[n] \hat{s}[n-m]\right\rbrace.
    \label{grad2}
\end{equation}
This particular gradient descent algorithm presents two stationary
situations (for which both Eq.~\ref{grad1} and Eq.~\ref{grad2} are
zero): $e[n] = 0$ and $\hat{y}[n]=0$, being the last one equivalent to
the condition $\hat{L}[m] = 0$.  As the first situation is clearly a
local minimum of the continuous cost function $\mathcal{E}$, the
second one must be a local maximum. As a consequence, always the
initial condition satisfies $\hat{L}[m] \neq 0$, the algorithm will
converges to $\mathcal{E}=0$. However, this condition does not
necessarily imply that the estimation is exact, i.e.
$\hat{L}[m] = L[m]$ or $\hat{O}[m] = O[m]$, and further analysis is
required in order to study the stationary state of the estimation
coefficients.

An useful investigation, similar to that done to obtain
Eq.~\ref{lms_convergence}, is the frequency-domain expression of the
measurement signal. In Appendix~\ref{frequencyconver} we show that the
stationary state $e[n] = 0$, in the case of a nonlinear observation
channel, involves the frequency-domain relation
\begin{equation}
  \breve{O}(\Omega^q_s)\prod_{i=1}^q \breve{L}(\Omega_i) = \tilde{O}(\Omega^q_s)\prod_{i=1}^q \tilde{L}(\Omega_i) \quad \forall\quad\lbrace q | a_q\neq0 \rbrace,
    \label{fund01}
\end{equation}
where $\Omega^q_s = \sum_{i=1}^{q} \Omega_i$. We observe that, unlike
in the case of Eq.~\ref{lms_convergence}, a low-bandwidth observation
channel does not preclude the information about high-frequency
components of the estimated channel. This is because for any $q>1$,
there is always a combination of high frequencies $\Omega_i$ such that
$\Omega^q_s$ lies within the observation channel bandwidth, leading to
accessible information about the components $\tilde{L}(\Omega_i)$. As
an alternative interpretation, the nonlinearity always translates
information of high-frequency components of $y[n]$ to the
low-frequency spectrum, allowing this information to surpass the
low-bandwidth observation channel. However, in what follows, we
provide a formal justification of these approaches, showing that the
estimation of the channel is possible and it does not depends on the
observation channel bandwidth.

In order to find a more direct relation between $\hat{L}[m]$ and
$L[m]$ we rewrite Eq.~\ref{fund01} as
\begin{equation}
    A(\Omega^q_s)\prod_{i=1}^q B(\Omega_i) = 1,
    \label{fund02}
\end{equation}
where $A(\Omega) = \breve{O}(\Omega)/\tilde{O}(\Omega)$ and
$B(\Omega) = \breve{L}(\Omega)/\tilde{L}(\Omega)$. By derivation of
Eq.~\ref{fund02} with respect to $\Omega_j$, with
$j\in\lbrace 1,2,..,q \rbrace$, and dividing the result by the left
member of Eq.~\ref{fund02} we obtain
\begin{equation}
    \frac{B'(\Omega_j)}{B(\Omega_j)} = -\frac{A'(\Omega^q_s)}{A(\Omega^q_s)}.
\end{equation}
As the left member only depends on $\Omega_j$, while the right member
depends of all the variables $\Omega_i$, the only way to satisfy this
relation is
\begin{equation}
    \frac{B'(\Omega_j)}{B(\Omega_j)} = -\frac{A'(\Omega^q_s)}{A(\Omega^q_s)} = c,
\end{equation}
where $c$ is an arbitrary complex constant. The solution of this
differential equation leads to
\begin{equation}
    \left \lbrace \begin{array}{l} B(\Omega) = e^{c\Omega + c_B} \\ A(\Omega) = e^{-c\Omega + c_A},\end{array} \right.
    \label{solution01}
\end{equation}
where $c_B$ and $c_A$ are also arbitrary complex numbers. By replacing
the definitions of $A(\Omega)$ and $B(\Omega)$ in Eq.~\ref{solution01}
we obtain
\begin{equation}
  \left \lbrace \begin{array}{l} \breve{L}(\Omega) = \tilde{L}(\Omega)e^{c\Omega + c_B} \\ \breve{O}(\Omega) = \tilde{O}(\Omega)e^{-c\Omega + c_A}.\end{array} \right.
    \label{solution02}
\end{equation}
Taking into account that $\breve{L}(\Omega)$ and $\breve{O}(\Omega)$
are Fourier transforms of real sequences, they must satisfy
$\breve{L}(\Omega) = \breve{L}^*(-\Omega)$ and
$\breve{O}(\Omega) = \breve{O}^*(-\Omega)$. Consequently, $c$ is shown
to be an imaginary number, i.e. $c = i\tau$, and $c_A$ and $c_B$ are
shown to be real numbers. Thus we have
\begin{equation}
  \left \lbrace \begin{array}{l} \breve{L}(\Omega) = \alpha_B\tilde{L}(\Omega)e^{i\tau \Omega} \\ \breve{O}(\Omega) = \alpha_A\tilde{O}(\Omega)e^{-i\tau \Omega},\end{array} \right.
    \label{solution03}
\end{equation}
with $\alpha_B = e^{c_B}$ and $\alpha_A = e^{c_A}$. On the other hand, by replacing Eq.~\ref{solution03} in Eq.~\ref{fund01} we find that
\begin{equation}
  \alpha_A(\alpha_B)^q= 1 \quad \forall\quad\lbrace q | a_q\neq0 \rbrace.
  \label{alphaAB}
\end{equation}
Thus, if the nonlinear function has at least two coefficients
satisfying $a_q\neq 0$, one even and other odd, then necessarily
$\alpha_A=\alpha_B = 1$. Otherwise, if the nonlinear function poses
only one nonzero coefficient, $\alpha_A$ and $\alpha_B$ remain unknown
values satisfying Eq.~\ref{alphaAB}. In what follows we assume, for
simplicity, that $\alpha_A=\alpha_B = 1$.

By applying the inverse Fourier transform of Eq.~\ref{solution03} we
obtain
\begin{equation}
  \left \lbrace \begin{array}{l} \hat{L}[m] = L(t+\tau T_s)|_{t=mT_s} \\
                  \hat{O}[m] = O(t-\tau T_s)|_{t=mT_s},\end{array} \right.
  \label{solution04}
\end{equation}
where $1/T_s$ is the sampling frequency, while $L(t)$ and $O(t)$ are
continuous-time versions of $L[m]$ and $O[m]$, respectively, defined
as
\begin{equation}
  \left \lbrace \begin{array}{l} L(t) = \sum_n L[n] \mathrm{sinc}\{ \pi(t-nT_s)/T_s \} \\ O(t) = \sum_n O[n] \mathrm{sinc}\{\pi(t-nT_s)/T_s\},\end{array} \right.
\end{equation}
with $\mathrm{sinc}(x)=\frac{\sin(x)}{x}$. In other words, the
estimation of the channel is a temporal-shifted version of the
oversampled response of the actual channel response. It must to be
noted that this conclusion is not affected by the bandwidth of the
observation channel, suggesting that this estimation can be performed
even for an arbitrarily ultra-low bandwidth $\tilde{O}(\Omega)$. In
addition, if both channels are assumed to be causal ($L[m<0] = 0$ and
$O[m<0]= 0$) and not delayed ($L[0] \neq 0$ and $O[0] \neq 0$), the
only possibility to satisfy Eq.~\ref{solution04} is $\tau = 0$, being
the estimation completely exact.

\section{Numerical results}
In order to show an example of application of the proposed algorithm,
we perform numerical simulations of the system shown in
Fig.~\ref{fig03}, for a particular case. The estimated channel is
given by the arbitrary response
\begin{equation}
  L[m] = \frac{1}{N_L}\left(\frac{1}{4} + \mathrm{sin}(rm) + \mathrm{sin}(3rm/2) \right)e^{-25 m/N}, 
\end{equation}
with $1\leq m \leq N$, $r = 0.4$ and $N = 256$. $N_L$ is a normalization factor making $\sum_m L^2[m] = 1$. The observation channel is given by
\begin{equation}
  O[m] = \frac{1}{N_O}\left( \mathrm{sinc}(Bm) +\frac{1}{2}e^{-2Bm}\right),
\end{equation}
where $B$ is associated to the observation channel bandwidth and takes
three different values, $B = \{1/8,1/32,1/128\}$, and
$N_O$ is the normalization factor ensuring $\sum_m O^2[m] = 1$. The nonlinear function is described by
the coefficients $a_2 = 1$ and $a_3 = -2$; the rest of the
coefficients are assumed to be zero. The input sequence is obtained as
\begin{equation}
    x[n] = \frac{1}{N}\sum_{m=-N/2}^{N/2-1} x_0[n-m] \mathrm{sinc}(m/3),
    \label{input}
\end{equation}
where $x_0[m]$ is an stochastic variable with normal distribution
$\mathcal{N}(0,10)$. This definition ensures the oversampling of the
input sequence (Eq.~\ref{oversam}), i.e. $\tilde{x}(\Omega) \approx 0$
$\forall$ $|\Omega|>\pi/Q = \pi/3$. The estimation channels are
initialized as
\begin{equation}
  \hat{L}[m] = \hat{O}[m] = \delta_{m,0},
\end{equation}
where $\delta_{m,n}$ is the Kronecker delta. The adaptation step of
the gradient descent algorithm is set to $\beta = 0.1$.

\begin{figure}[h]
    \centering
    \begin{tikzpicture}
    \begin{axis}[ grid,ylabel near ticks, height=7.5cm,width=8.75cm,xlabel={$n$}, ylabel={$\mathcal{E}_n$ [dB]},xmin=1,xmax=1000000, xtick={200000,400000,600000,800000},
    xticklabels={2,4,6,8}]
    \addplot[line width = 1,black] file {Figs/E_8.dat};
    \addplot[line width = 1,blue,dashed] file {Figs/E_32.dat};
    \addplot[line width = 1,black!50!blue,densely dotted,smooth] file {Figs/E_128.dat};
    \legend{$B=1/8$,$B=1/32$,$B=1/128$}
    \end{axis}
    \end{tikzpicture}
    \caption{Evolution of the instantaneous mean error square $\mathcal{E}_n$ for three
      different bandwidths of the observation channel. These results
      suggest that the convergence of the algorithm does not depend on the observation channel bandwidth.}
\label{fig04}
\end{figure}
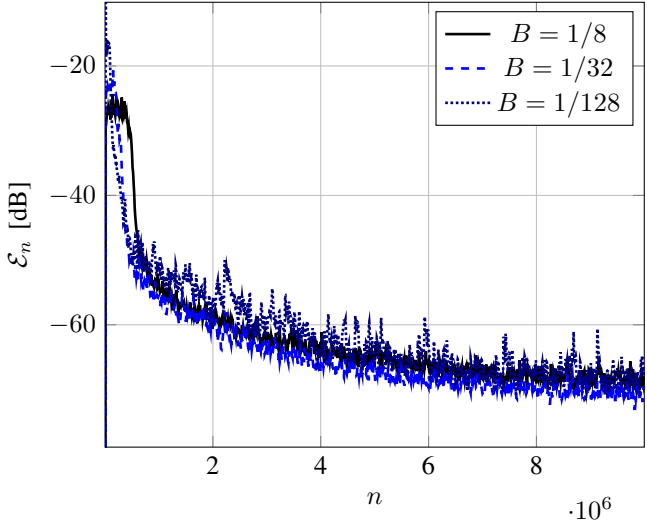

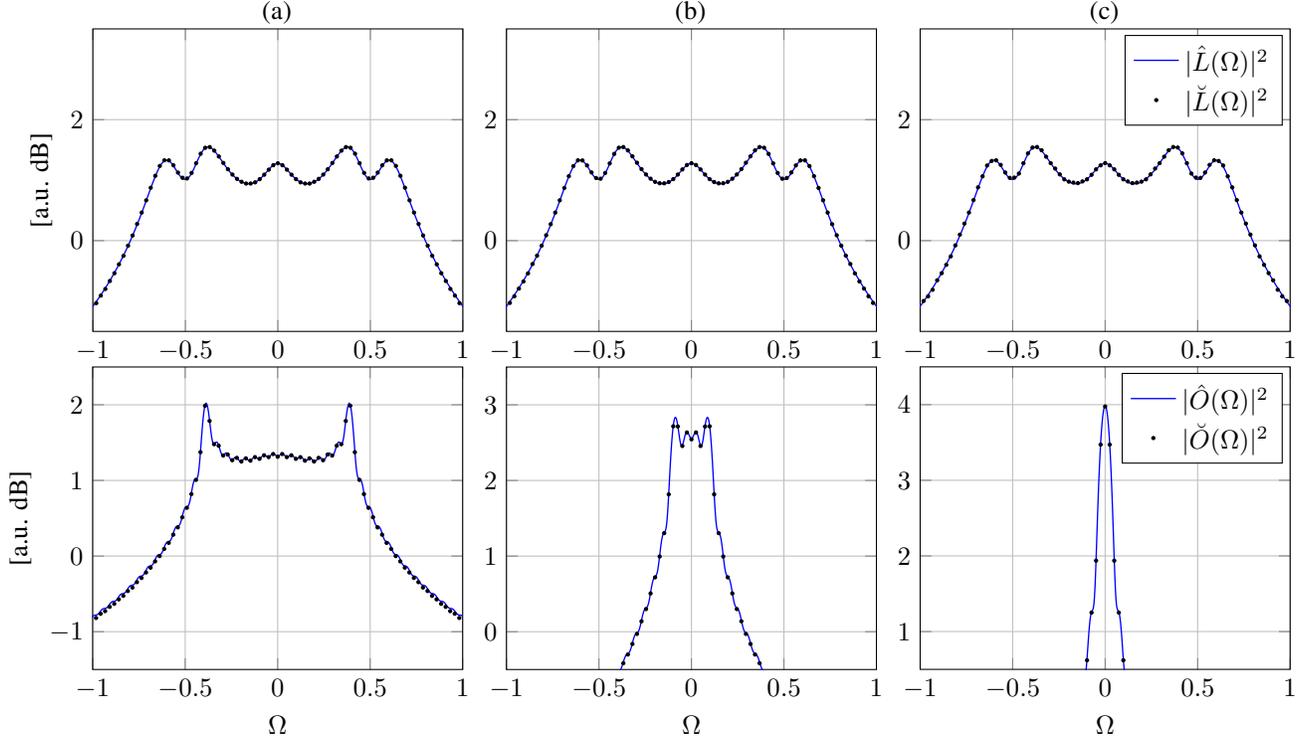
\begin{figure*}
    \centering
    \begin{tikzpicture}
    \node at (2.45,4.25) {(a)};
    \begin{axis}[ylabel={[a.u. dB]},grid,xmin=-1,xmax=1,ymin=-1.5,ymax=3.5,width=6.5cm,mark size = 0.6pt]
    \addplot[line width = 0.5,blue] file {Figs/L.dat};
    \addplot[only marks] file {Figs/A_8.dat};
    \end{axis}
    
    \tikzset{shift={(0,-4.5)}}
    
    \begin{axis}[ylabel={[a.u. dB]},xlabel={$\Omega$},grid,xmin=-1,xmax=1,ymin=-1.5,ymax=2.5,width=6.5cm,mark size = 0.6pt]
    \addplot[line width = 0.5,blue] file {Figs/O_8.dat};
    \addplot[only marks] file {Figs/B_8.dat};
    \end{axis}
    \tikzset{shift={(0,4.5)}}
    
    \tikzset{shift={(5.5,0)}}
    \node at (2.45,4.25) {(b)};
    \begin{axis}[grid,xmin=-1,xmax=1,ymin=-1.5,ymax=3.5,width=6.5cm,mark size = 0.6pt]
    \addplot[line width = 0.5,blue] file {Figs/L.dat};
    \addplot[only marks] file {Figs/A_32.dat};
    \end{axis}
    
    \tikzset{shift={(0,-4.5)}}
    \begin{axis}[xlabel={$\Omega$},grid,xmin=-1,xmax=1,ymin=-0.5,ymax=3.5,width=6.5cm,mark size = 0.6pt]
    \addplot[line width = 0.5,blue] file {Figs/O_32.dat};
    \addplot[only marks] file {Figs/B_32.dat};
    \end{axis}
    \tikzset{shift={(0,4.5)}}

    \tikzset{shift={(5.5,0)}}
    \node at (2.45,4.25) {(c)};
    \begin{axis}[grid,xmin=-1,xmax=1,ymin=-1.5,ymax=3.5,width=6.5cm,mark size = 0.6pt]
    \addplot[line width = 0.5,blue] file {Figs/L.dat};
    \addplot[only marks] file {Figs/A_128.dat};
    \legend{$|\hat{L}(\Omega)|^2$,$|\breve{L}(\Omega)|^2$}
    \end{axis}
    
    \tikzset{shift={(0,-4.5)}}
    \begin{axis}[xlabel={$\Omega$},grid,xmin=-1,xmax=1,ymin=0.5,ymax=4.5,width=6.5cm,mark size = 0.6pt]
    \addplot[line width = 0.5,blue] file {Figs/O_128.dat};
    \addplot[only marks] file {Figs/B_128.dat};
    \legend{$|\hat{O}(\Omega)|^2$,$|\breve{O}(\Omega)|^2$}
    \end{axis}
    \tikzset{shift={(0,4.5)}}
    
    \end{tikzpicture}
    \caption{Estimation of the channel of interest (top) and the
      observation channel (bottom), obtained with the proposed LMS
      algorithm, after $n = 1000000$ steps. Three different bandwidth
      of the observation channel are considered; $B = 1/8$ (a),
      $B = 1/32$ (b) and $B=1/128$ (c). The performance of the
      algorithm is not affected by the extremely low bandwidth of
      $\tilde{O}(\Omega)$.}
    \label{fig05}
\end{figure*}

Figure~\ref{fig04} shows the evolution of the mean error square for
the three cases simulated, corresponding to different values of
channel observation bandwidth. On the other hand, Fig.~\ref{fig05}
displays the obtained estimations in comparison with the actual
channel frequency-responses. It must be noted that the system
identification is accurate even for the extreme case $B = 1/128$, for
which the observation channel bandwidth is ultra-low with respect to
the estimated channel bandwidth. This is a clear sign that the
convergence of the algorithm does not depend on the the $O[m]$
bandwidth.

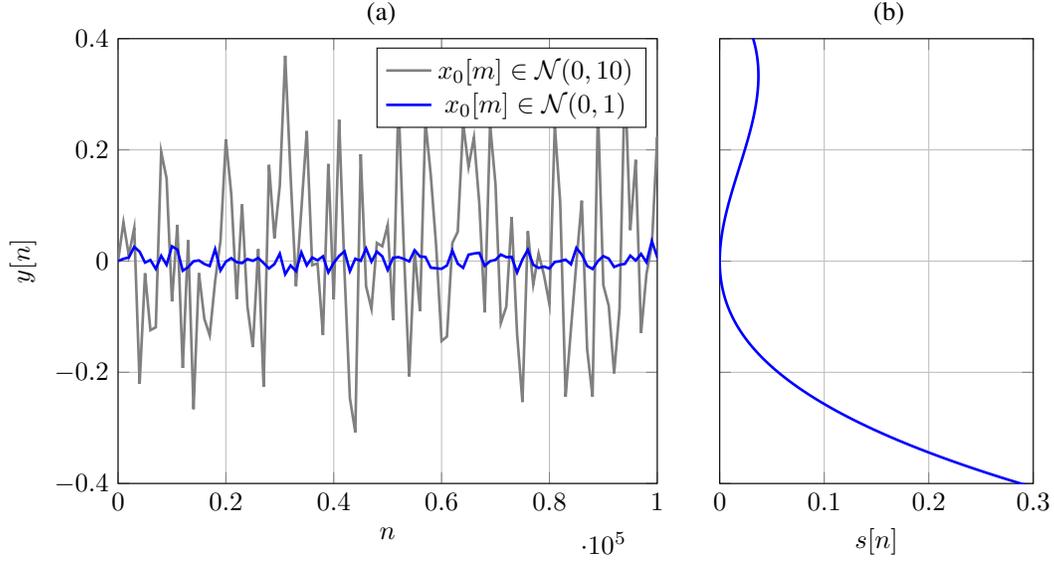
\begin{figure*}[h]
    \centering
    \begin{tikzpicture}
    \begin{axis}[grid,ylabel near ticks, height=7.5cm,width=8.75cm,xlabel={$n$}, ylabel={$y[n]$},xmin=0,xmax=100000,ymin=-0.4,ymax=0.4]
    \addplot[line width = 1,gray] file {Figs/Y_big.dat};
    \addplot[line width = 1,blue] file {Figs/Y_small.dat};
    \legend{{$x_0[m]\in\mathcal{N}(0,10)$},{$x_0[m]\in\mathcal{N}(0,1)$}};
    \end{axis}
    
    \tikzset{shift={(8,0)}}
    
    \begin{axis}[grid,ylabel near ticks, height=7.5cm,width=5.75cm,xlabel={$s[n]$},xmin=0,xmax=0.3,ymin=-0.4,ymax=0.4, yticklabels={}]
    \addplot[line width = 1,blue] file {Figs/NLfun.dat};
    \end{axis}
    
    \node at (-4.5,6.25) {(a)};
    \node at (2.25,6.25) {(b)};
    \end{tikzpicture}
    \caption{(a) Vector sample of the output sequence for two different input sequences given by Eq.~\ref{input}. (b) Memoryless nonlinear function modeled with coefficients $a_2 = 1$ and $a_3 = -2$. In the case odf the low-power output sequence, $s[n]$ can be clearly obtained with an effective quadratic nonlinear function, neglecting the third-order term.}
\label{fig06}
\end{figure*}

\begin{figure}
    \centering
    \begin{tikzpicture}
    \begin{axis}[ylabel={[a.u. dB]},grid,xmin=-1,xmax=1,ymin=-1.5,ymax=3.5,width=8.5cm,mark size = 0.6pt]
    \addplot[line width = 0.5,blue] file {Figs/L.dat};
    \addplot[only marks] file {Figs/A2_32.dat};
    \legend{$|\hat{L}(\Omega)|^2$,$|\breve{L}(\Omega)|^2$}
    \end{axis}
    
    \tikzset{shift={(0,-6.5)}}
    
    \begin{axis}[ylabel={[a.u. dB]},xlabel={$\Omega$},grid,xmin=-1,xmax=1,ymin=-1.5,ymax=3.5,width=8.5cm,mark size = 0.6pt]
    \addplot[line width = 0.5,blue] file {Figs/O_8.dat};
    \addplot[only marks] file {Figs/B2_32.dat};
    \legend{$|\hat{O}(\Omega)|^2$,$|\breve{O}(\Omega)|^2$}
    \end{axis}
    \end{tikzpicture}
    \caption{Estimation of the channel of interest (top) and the observation channel (bottom) with $B = 1/8$ with a lower-power input sequence, leading to a single quadratic term effective nonlinear function. The estimation differs of the actual channel by two unknown scaling factors. In coherent optical devices, this drawback can be compensated by the simple measuring of the mean output power.}
    \label{fig07}
\end{figure}
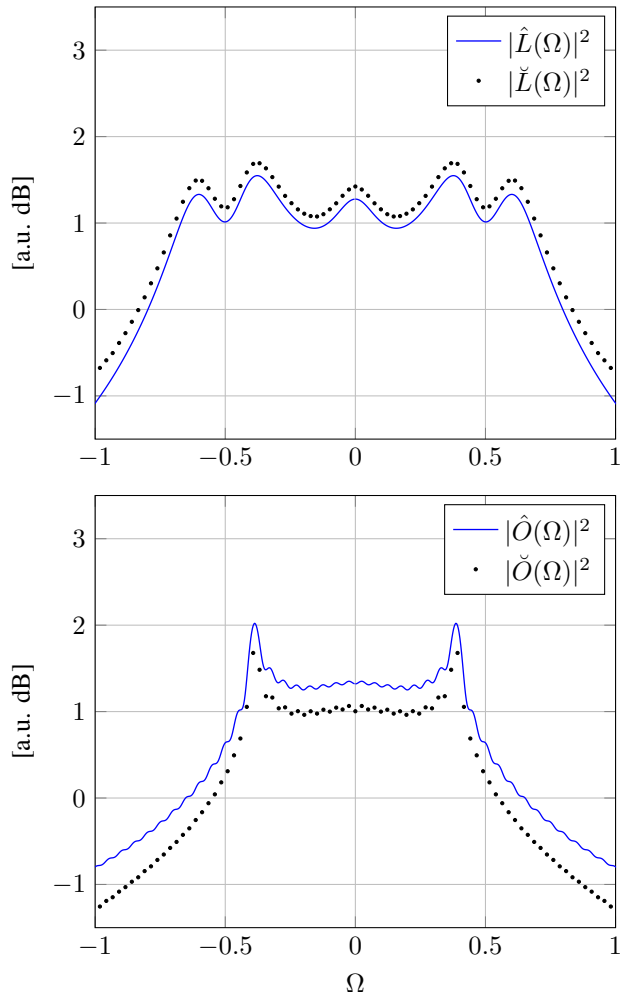

The nonlinear function of the previous example has one odd coefficient and another even, allowing for an exact estimation of the system. However, as the input-sequence power is reduced, the relevance of the third-order term is also diminished, having an effective quadratic feedback with only one coefficient, $a_2$. This situation is illustrated in Fig.~\ref{fig06}, where we show a vector sample of the output sequence $y[n]$ compared with the nonlinear function. We observe that the signal excursion is enough to provide a cubic behavior of the relation between $s[n]$ and $y[n]$. On the other hand, when the output sequence is obtained with a lower-power input sequence, the nonlinear block works as a single-term quadratic function. Consequently, the estimation of the channels is given by Eq.~\ref{solution03}, where $\alpha_A$ and $\alpha_B$ are two unknown scaling factors satisfying $\alpha_A\alpha^2_B = 1$. We repeat the simulations for the case $B=1/8$ but reducing the input-sequence power. As shown in Fig.~\ref{fig07}, the shape of the frequency responses is correctly estimated, but their scale do not fit with that of the actual channels. However, this drawback can be circumvent if an additional constrain is added to the algorithm. For instance, in coherent optical devices where the feedback channel is implemented with a common photodiode, the nonlinear function is naturally modeled with a simple quadratic term, leading to this scaling problem. By the simple measurement of the mean optical power of the output ($E\{y^2[n]\}$), the estimation $\hat{L}[m]$ can be re-scaled to obtain $E\{\hat{y}^2[n]\} = E\{y^2[n]\}$, thus obtaining the exact factor correction.

\section{Conclusions}
We shown that the system identification of a linear channel with an
unknown-response observation channel is possible as long as a
nonlinear memoryless function, uncoupling both systems, is
introduced. In particular, the estimation of the channel is a scaled
and temporal-shifted version of the actual channel, being the scaling
and shifting factors two unknown coefficients. However, if the
nonlinear function, expressed as a polynomial sum, has at least one
term of even order and other of odd order, the scaling factor is
necessarily the unity. In addition, if the channels are assumed to be
causal and not delayed, the shifting factor is zero, allowing for a
completely faithful estimation of both systems. Moreover, unlike in
the standard scheme of a chain of linear systems, the estimation does
not depend on the measurement system bandwidth. Actually, by means of
a few illustrating numerical examples, the proposed LMS algorithm was
shown to be effective even for ultra-low bandwidth observation
channels, in comparison with the bandwidth of the channel of interest. These results suggest that this scheme can be implemented with low-cost measurement device. A clear example can be found in the coherent optical devices, where the observation channel can be reduced to a low-cost standard photodiode.

\begin{small}
\appendices
\section{Stationaty State of the LMS Algorithm, Including the
  Estimation of the Observation Channel}\label{lms_problem}
We first derive a frequency-domain expression for the measurement
sequence $z[n]$. By replacing Eq.~\ref{firsteq} into
Eq.~\ref{measuresignal} we obtain
\begin{equation}
    z[n] = \sum_{m,r} L[m]O[r]x[n-m-r].
\end{equation}
Using the inverse Fourier transform we find that
\begin{multline}
  \frac{1}{2\pi} \int_{-\pi}^{\pi} \tilde{z}(\Omega) e^{\dot{\j}\Omega
    n} \,d\Omega =\\ \frac{1}{(2\pi)^3}\sum_{m,r}
  \iiint_{-\pi}^{\pi}\tilde{L}(\Omega_1)
  \tilde{O}(\Omega_2)\tilde{x}(\Omega) \times \\ e^{\dot{\j}\Omega_1 m
    + \dot{\j}\Omega_2 r + \dot{\j}\Omega (n-m-r)} \,d\Omega_1
  \,d\Omega_2\,d\Omega.
\end{multline}
We sum the variables $m$ and $r$, taking into account the relation
\begin{equation}
    \sum_m e^{\dot{\j}\Omega m} = 2\pi \sum_u \delta(\Omega + 2\pi u),
    \label{deltadirac}
\end{equation}
where $\delta(.)$ is the Dirac delta function, to obtain
\begin{equation}
  \frac{1}{2\pi} \int_{-\pi}^{\pi} \tilde{z}(\Omega) e^{\dot{\j}\Omega n} \,d\Omega = \frac{1}{2\pi} \int_{-\pi}^{\pi}\tilde{L}(\Omega) \tilde{O}(\Omega)\tilde{x}(\Omega) e^{\dot{\j}\Omega n} \,d\Omega.
\end{equation}
Thus we find that
\begin{equation}
  \tilde{z}(\Omega) = \tilde{L}(\Omega) \tilde{O}(\Omega)\tilde{x}(\Omega).
  \label{AppA01}
\end{equation}
A similar procedure can be performed to obtain the Fourier transform
of the estimated measure signal $\hat{z}[n]$, that is
\begin{equation}
  \breve{z}(\Omega) = \breve{L}(\Omega) \breve{O}(\Omega)\tilde{x}(\Omega).
  \label{AppA02}
\end{equation}
In the stationary state of the LMS algorithm the error signal is zero,
$e[n] = \hat{z}[n] - z[n] = 0$. As a consequence, Eq.~\ref{AppA01}
must be equal to Eq.~\ref{AppA02}, proving the relation
\begin{equation}
  \breve{L}(\Omega) \breve{O}(\Omega) = \tilde{L}(\Omega) \tilde{O}(\Omega).
\end{equation}

\section{Derivatives of the Gradient Descent Algorithm}\label{derivatives}
We calculate the derivatives of the mean error square,
$\mathcal{E}_n = \lbrace e^2[n] \rbrace$, where
$e[n] = \hat{z}[n] - z[n]$. On one hand, its derivative with respect
to the estimation of the observation channel coefficients is given by
\begin{multline}
  \frac{\partial \mathcal{E}_n}{\partial \hat{O}[m]} = 2\left\lbrace
    e[n] \frac{\partial e[n]}{\partial \hat{O}[m]}\right\rbrace =
  2\left\lbrace e[n] \frac{\partial \hat{z}[n]}{\partial
      \hat{O}[m]}\right\rbrace = \\ 2\left\lbrace e[n]
    \frac{\partial}{\partial \hat{O}[m]} \sum_r
    \hat{O}[r]\hat{s}[n-r]\right\rbrace = 2\left\lbrace e[n]
    \hat{s}[n-m]\right\rbrace.
\end{multline}
On the other hand, the derivative with respect to the coefficients of
the estimation of the channel is given by
\begin{multline}
  \frac{\partial \mathcal{E}_n}{\partial \hat{L}[m]} = 2\left\lbrace
    e[n] \frac{\partial e[n]}{\partial \hat{L}[m]}\right\rbrace =
  2\left\lbrace e[n] \frac{\partial \hat{z}[n]}{\partial
      \hat{L}[m]}\right\rbrace =\\ 2\left\lbrace e[n] \sum_r
    \hat{O}[r]\frac{\partial\hat{s}[n-r]}{\partial\hat{L}[m]}\right\rbrace
  =\\ 2\left\lbrace e[n] \sum_r \hat{O}[r]\sum_{q=1}^Q
    a_q\frac{\partial \hat{y}^q[n-r]}{\partial\hat{L}[m]}\right\rbrace
  = \\2\left\lbrace e[n] \sum_r \hat{O}[r]\sum_{q=1}^Q a_q q
    \hat{y}^{q-1}[n-r]\frac{\partial
      \hat{y}[n-r]}{\partial\hat{L}[m]}\right\rbrace =\\ 2\left\lbrace
    e[n] \sum_r \hat{O}[r]\sum_{q=1}^Q a_q q \hat{y}^{q-1}[n-r]
    \frac{\partial }{\partial\hat{L}[m]}\sum_u \hat{L}[u]
    x[n-r-u]\right\rbrace = \\ 2\left\lbrace e[n] \sum_r
    \hat{O}[r]\sum_{q=1}^Q a_q q \hat{y}^{q-1}[n-r]
    x[n-r-m]\right\rbrace
\end{multline}

\section{Stationary state of the LMS Algorithm with a Nonlinear
  Observation Channel}\label{frequencyconver}
We start from the derivation of a complete expression of the
measurement sequence $z[n]$. By following the scheme of
Fig.~\ref{fig03} and the definition of each block we obtain
\begin{multline}
  z[n] = \sum_m O[m] s[n-m] = \sum_m O[m]\sum_{q=1}^Qa_qy^q[n-m] =\\
  \sum_m O[m]\sum_{q=1}^Qa_q\left[\sum_r L[r]x[n-m-r]\right]^q.
\end{multline}
By using the inverse Fourier transform of $L[m]$ and $x[m]$ we find
\begin{multline}
  z[n] = \sum_m O[m]\sum_{q=1}^Qa_q \times\\
  \left[\frac{1}{(2\pi)^2}\sum_r
    \iint_{-\pi}^{\pi}\tilde{L}(\Omega_1)\tilde{x}(\Omega_2)e^{\dot{\j}\Omega_1
      r + \dot{\j}\Omega_2(n-m-r)} \,d\Omega_1\,d\Omega_2\right]^q.
\end{multline}
By summing over $r$, and taking into account Eq.~\ref{deltadirac}, we
obtain
\begin{equation}
  z[n] = \sum_m O[m]\sum_{q=1}^Qa_q \left[\frac{1}{2\pi} \int_{-\pi}^{\pi}\tilde{L}(\Omega)\tilde{x}(\Omega)e^{ \dot{\j}\Omega(n-m)} \,d\Omega\right]^q.
\end{equation}
This equation can be rewritten in the more convenient way,
\begin{multline}
  z[n] = \sum_{q=1}^Qa_q \sum_m O[m]\frac{1}{(2\pi)^q}\times\\ \int
  ..\int_{-\pi}^{\pi} \left[ \prod_{i=1}^Q
    \tilde{L}(\Omega_i)\tilde{x}(\Omega_i)e^{
      \dot{\j}\Omega_i(n-m)}\right] \,d\Omega_1 ..\,d\Omega_q = \\
  \sum_{q=1}^Qa_q \sum_m O[m]\frac{1}{(2\pi)^q}\times\\ \int
  ..\int_{-\pi}^{\pi} e^{-\dot{\j}\Omega^q_s m}\left[ \prod_{i=1}^q
    \tilde{L}(\Omega_i)\tilde{x}(\Omega_i)e^{ \dot{\j}\Omega_i
      n}\right] \,d\Omega_1 ..\,d\Omega_q,
\end{multline}
where
\begin{equation}
    \Omega^q_s = \sum_{i=1}^q \Omega_i.
\end{equation}
By using the inverse Fourier transform of $O[m]$ we find that
\begin{multline}
  z[n] = \sum_{q=1}^Qa_q \frac{1}{(2\pi)^{q+1}}\sum_m\iint
  ..\int_{-\pi}^{\pi}\tilde{O}(\Omega)e^{\dot{\j}\Omega m} \times\\
  e^{-\dot{\j}\Omega^q_s m}\left[ \prod_{i=1}^q
    \tilde{L}(\Omega_i)\tilde{x}(\Omega_i)e^{ \dot{\j}\Omega_i
      n}\right] \,d\Omega_1 ..\,d\Omega_q\,d\Omega.
    \label{equa28}
\end{multline}
We assume the input signal to be oversampled in a factor $Q$, that is
\begin{equation}
  \tilde{x}(\Omega) = 0 \quad \forall \quad |\Omega|\geq \pi/Q.
    \label{oversam}
\end{equation}
Consequently, by summing Eq.~\ref{equa28} over $m$ we obtain
\begin{multline}
  z[n] = \sum_{q=1}^Qa_q \frac{1}{(2\pi)^{q}}\int
  ..\int_{-\pi}^{\pi}\tilde{O}(\Omega^q_s)e^{ \dot{\j}\Omega_s n}
  \times\\ \left[ \prod_{i=1}^q
    \tilde{L}(\Omega_i)\tilde{x}(\Omega_i)\right] \,d\Omega_1
  ..\,d\Omega_q.
    \label{measure}
\end{multline}
Analogously, we can prove that the estimated measure sequence,
$\hat{z}[n]$ can be expressed as
\begin{multline}
  \hat{z}[n] = \sum_{q=1}^Qa_q \frac{1}{(2\pi)^{q}}\int
  ..\int_{-\pi}^{\pi}\breve{O}(\Omega^q_s)e^{ \dot{\j}\Omega_s n}
  \times\\ \left[ \prod_{i=1}^q
    \breve{L}(\Omega_i)\tilde{x}(\Omega_i)\right] \,d\Omega_1
  ..\,d\Omega_q.
    \label{estimate}
\end{multline}
At the stationary state of the LMS algorithm, the measure sequence is
assumed to be equal to its estimation, i.e. $z[n] = \hat{z}[n]$. By
replacing Eqs.~\ref{measure}~and~\ref{estimate} in this equlality we
find that
\begin{equation}
  \breve{O}(\Omega^q_s)\prod_{i=1}^q \breve{L}(\Omega_i) = \tilde{O}(\Omega^q_s)\prod_{i=1}^q \tilde{L}(\Omega_i) \quad \forall \quad \lbrace q | a_q\neq0 \rbrace.
\end{equation}
\end{small}

\ifCLASSOPTIONcaptionsoff
  \newpage
\fi


\begin{thebibliography}{9}
\bibitem{pintelon2012system}
R. Pintelon and J.Schoukens, \emph{System identification: a frequency domain approach}, John Wiley and Sons, 2012.

\bibitem{keesman2011system}
K. J. Keesman, \emph{System identification: an introduction}, Springer Science and Bussiness Media, 2011.

\bibitem{widrow1985adaptive}
B. Widrow and S. D. Stearns, \emph{Adaptive Signal Processing}, Englewood Cliffs, 1985.

\bibitem{schoukens2019nonlinear}
J. Schouckens and L. Ljung, \emph{Nonlinear System Identification: A user-oriented road map an introduction}, IEEE Control Systems Magazine, 39, 28-99, 2019.

\end{thebibliography}
\end{document}